# Crystal Chemical Estimation of possible exceeding of $T_c$ 77 K in Diborides

**L.M. Volkova, [1] S.A. Polyshchuk, [1] S.A. Magarill [2], F.E. Herbeck [1]**

[1] Institute of Chemistry, Far Eastern Branch RAS, 690022 Vladivostok, Russia
[2] Institute of Inorganic Chemistry, Sib. Branch RAS, 630090 Novosibirsk, Russia.

## Abstract

We establish the empirical correlation between $T_c$ of diborides ($AB_2$) and crystal chemical parameters of simpler structural fragment – sandwich $A_2(B_2)$ where the superconductivity is appeared, like found in high-$T_c$ cuprate superconductors. Our results suggest that in the absence of vacancies in $B_2$ plane $T_c$ of diborides can be higher 77 K, the evaporation temperature of liquid nitrogen. We discuss critical crystal chemistry parameters controlling $T_c$ and ways to achieve higher transition temperatures in diborides based on this correlation.

## 1. INTRODUCTION

Attempts to exceed the temperature of transition to superconductivity ($T_c$) 39 K [1] in the $AB_2$ diborides row with the $AlB_2$ structure have until no success. However, the presence of some common crystal chemical characteristics in diborides and high-temperature superconductor cuprate (HTSC) cuprates allows to expect that possibility of raising $T_c$ in these compounds still exist. Intensive study of $MgB_2$ had shown that as in HTSC cuprates the layered nature of $MgB_2$ is caused anisotropic superconducting properties [2, 3]. It was shown that $T_c$ of $MgB_2$ either as in HTSC cuprates depends on the concentration of charge carriers [4, 5] and structural parameters [6, 7]. The similar ways of variation in concentration of charge carriers and structural parameters by isovalent and non-isovalent doping [7, 8-15] and by pressure [5, 16-23] also were considered. It was established that in diborides the $B_2$ planes contain such carriers of charge as holes (p) and play the same role as the $CuO_2$ in high-$T_c$ cuprates [4, 24, 25]. In $AB_2$ these planes there are between the planes of positive charged "ions" A, as the $CuO_2$ planes in perovskite layer $A_{n+1}(CuO_2)_n$. Note, that in the last this is reached for account of displacement or full removing the apical oxygen atoms from plane of A cations on consequence of Jan-Teller effect.

In [26] we have shown that the simplest structural fragment in HTSC cuprates is not one $CuO_2$ plane but $A_2(CuO_2)$ sandwich, in which the $CuO_2$ plane with charge carriers is situated between the planes of A-cations, and have installed the correlation of $T_c$ with critical crystal chemical parameters of this sandwich general for all phases of HTSC cuprates. It is possible to expect that in diborides $AB_2$ of such structural fragment is the $A_2(B_2)$ sandwich, where the network of B atoms with charge carriers is situated between the planes of A "cations".

In this work we consider a dependence of $T_c$ from crystal chemical parameters of $A_2(B_2)$ sandwich in diborides like found in HTSC cuprates and discuss the possibilities and ways to rise $T_c$ in this class of compounds.

## 2. METHOD

In high–$T_c$ cuprates $T_c$ dependence on crystal chemical parameters of the $A_2(CuO_2)$ fragment is more full expressed by combination of such values as distances $d$(Cu-Cu) in $CuO_2$ plane and ones $d(CuO_2\text{-}A)$ from $CuO_2$ plane to adjacent plane of A cations, which have an original sense and give also an information about the hole concentration, and on the size and charge of A-cations and doping atoms too. In [26] we established the empirical dependence of $T_c$ from the ratio ($J$) of distances between Cu atoms along diagonal direction of $CuO_2$-plane to sum "effective" distances ($D_1+D_2$) from $CuO_2$ plane to of two adjacent planes of A cations in $A_{n+1}(CuO_2)_n$ layer, taking into account the charge and the size of A-cations and doping atoms ($J = d(\text{Cu-Cu})/(D_1+D_2)$).

For plotting the dependence of $T_c(J)$ in $AB_2$ where

$$J = d(B-B)/(D_1 + D_2) \qquad (1)$$

we have chosen critical parameters of $A_2(B_2)$ fragment, like found in high–$T_c$ cuprates, such as:

1. The $d$(B-B) distances between B atoms situated at the ends of diagonal of hexagons in $B_2$ plane ($d(B-B) = 2a/\sqrt{3}$), i.e., at the maximal possible distance from one another. It is supposed, that the diagonal pairing and oscillation of holes on B atoms are energetically more favourable than other configurations [27];

2. “Effective” distances $D_1$ and $D_2$ from $B_2$ plane to surface of two adjacent planes of A “cations”:

$$D = S[d(B_2 - A) - R_A(Z_A/2)] \quad (2)$$

where $d(B_2\text{-}A)$ is the distance from $B_2$ plane to plane of A “cations”; $R_A$ is radius of A “cation”, which content is maximum; $Z_A/2$ – undimensional coefficient to take into account of the electric field of the A “cation” charge (it is the ratio of charge A “cation” to charge of Mg cation), $S$ is deviation coefficient of parameters of doping cations from parameters of A-cation that forms the plane:

$$S \geq 1,\ S = \overline{R(Z/2)}/R_A(Z_A/2) \text{ or } S = R_A(Z_A/2)/\overline{R(Z/2)} \quad (3)$$

Here $\overline{R(Z/2)}$ is generalized value, characterized the plane of A-cations:

$$\overline{R(Z/2)} = m_1 R_{A_1}(Z_{A_1}/2) + \ldots m_n R_{A_n}(Z_{A_n}/2) \quad (4)$$

where $m_n$ is content of $A_n$ “cation” in plane, $R_{A_n}$ is radius, $Z_{A_n}/2$ – undimensional coefficient to take into account of the electric field of the $A_n$ “cation” charge.

On value $J$ an enormous influence renders a size and a charge (valence states) of A “cations”. To value these parameters in intermetal compounds is not easy. By interpretation the result on diborides practically in all works the Pauling “crystal” ion radii are used [28]. However, this system reflects unadequately the changing of a lattice parameter $c$ by changing a radius of A “cations”. So, lattice parameter $c$ of $AlB_2$ (3.26 Å) is noticeable less than $c$ parameter of $MgB_2$ (3.52 Å) [29]. Pauling radii (*RP*) Al (0.75 Å), opposite, exceeds a radius of Mg (0.65 Å). Proceeding from approximately linear correlations between $c$ parameters of diborides and A radius the Shannon crystal radius (*RSh*) is more suitable. In Fig. 1, lattice parameters $c$ are plotted as a function of Shannon (a) and Pauling (b) radii of A “cations”. However, $J$ was calculated by using both Pauling and Shannon radii systems with standard coordination 6 and valent state equal group number in Periodic system, excluding Mn, Cr, Ru and Os. The valent state of Mn and Cr was accepted as 4+ and 3+, accordingly, but Ru and Os as 6+. Moreover, Ru radius was accepted such as for Os, as there is no data about Ru radius.

Besides, for the plotting of $T_c(J)$ it is necessary to know $T_c$ of sample and its full structural data (atomic coordinates and occupancy positions) obtained by X-ray and neutron diffraction. It is not sufficiently the only data about unit cell parameters, as the vacancies in A plane can change a $J$ value and, accordingly, $T_c$ but the vacancies in $B_2$ plane, as in $CuO_2$ plane of HTSC cuprates, reduce $T_c$ or suppress completely a superconductivity. Besides, the impurities in A-planes can be distributed irregularly along the *c* axis, as that was shown in [13] for $Mg_{1-x}Al_xB_2$ by high-resolution transmission-electron microscopy investigation. This will certainly result in a change and inequality of "effective" distances $D_1$ and $D_2$, accordingly, and to change J and superconducting characteristics. There isn't now such data on doping borides.

Realistically it is installed only the following:

(1) Crystal structure of diborides, lattice parameters and $T_c$ of polycrystalline and tin films of many $MgB_2$ samples [1, 5, 7, 31-35], and the structural data obtained by X-ray diffraction analysis on single crystal of only one $MgB_2$ sample [36], as well as the lattice parameters and the absence of superconductivity in $AlB_2$[10, 12, 13];

(2) Reducing of $MgB_2$ $T_c$ with the growing of pressure (P). Moreover the spread of d$T_c$/dP reported by different groups [5, 16-23] is high and is explained in [21] by various in the sample stoichiometry.

(3) The substitution of Al for Mg in $MgB_2$ decreases the $T_c$ and leads to the loss of superconductivity in $Mg_{1-x}Al_xB_2$ with the growing $x$ [7, 10-13].

It has been also found the superconducting transition at $T_c$=9.5 K for $TaB_2$ [37, 38] and no superconducting for $TiB_2$, $HfB_2$, $ZrB_2$, $VB_2$ and $NbB_2$. Although this result contradicts the data reported in [39], by which $ZrB_2$ is superconducting with $T_c$=5.5 K, and $TaB_2$ and $NbB_2$ are not. Further to this we have the experimental and theoretical data on superconductivity and structural properties of $Mg_{1-x}A_xB_2$ (A = Al, Zn, Ca and Na [7-13, 40] and only structural parameters of the other members of the $AB_2$ family (A = Ru, Os, Cr, Mn, Mo, W, V, Nb, Ta, Ti, Zr, Hf, Sc, Y, Cu, Ag, Au Lu, Pu and U) [29, 41-62] on superconducting properties of which not yet made final conclusions.

In this connection we have calculated two variants of $T_c(J)$ correlation: with only real data for compounds $MgB_2$ (single crystal) [36], $AlB_2$ [29, 57] and disputable data on $TaB_2$ [37] or $ZrB_2$ [39]. The both systems of radii (*RSh* and *RP*) were used by this calculation. As a result we have got two correlations for the variant-I ($T_c^{Ta,RSh}(J_{RSh})$ and $T_c^{Ta.RP}(J_{RP})$ (Fig. 2 a)) and two correlations for variant-II ($T_c^{Zr,RSh}(J_{RSh})$ and $T_c^{Zr,RP}(J_{RP})$ (Fig. 2 b)). The equations of second degree polynomial give the best approximation of the correlations:

$T_c^{Ta,RSh} = -6.05213 J_{RSh}^2 - 16.9383 J_{RSh} + 95.324$, there midpoint $J_0$ = -1.30395, $T_{c,\max}$=107; (5)

$T_c^{Ta,RP} = -9.93835 J_{RP}^2 - 68.1881 J_{RP} + 173.101$, there midpoint $J_0$ = -3.38595, $T_{c,\max}$=290; (6)

$T_c^{Zr,RSh} = 1.21806 J_{RSh}^2 - 51.7357 J_{RSh} + 135.714$, there midpoint $J_0$ = 20.8830, $T_{c,\min}$= -413; (7)

$T_c^{Zr,RP} = 11.0727 J_{RP}^2 - 143.304 J_{RP} + 239.537$, there midpoint $J_0$ = 6.3148 $T_{c,\,min}$= -224; (8)

Then, for the estimation of validity of these two correlation variants $J_{RSh}$ and $J_{RP}$ by formula (1) and corresponding them $T_c$ on equations (5-8) for 100 considered diborides were calculated. It was found that parabolic dependencies $T_c^{Ta,RSh}(J_{RSh})$ and $T_c^{Ta.RP}(J_{RP})$ of Eq. 5 and 6 obtained on the data for $MgB_2$, $AlB_2$ and $TaB_2$ estimate adequately a trend of changing $T_c$ of $MgB_2$ with the growing of pressure, and $Mg_{1-x}Al_xB_2$ with *x* increase, and confirm also an absence of superconductivity in $ZrB_2$ [37] (Table I). However, a velocity of linear falling $T_c$ ($dT_c/dP$ and $dT_c/dx$) calculated on these correlations is more below than experimental [7, 10-13]. Moreover, according the data, obtained by these correlations, a very small changing of lattice paramers can result in the lost of superconductivity in $TaB_2$. So, $TaB_2$ (Table I, sample N 28) in which the authors [39] did not find a superconductivity, by calculations on correlation $T_c^{Ta.RP}(J_{RP})$ of Eq. 6 also is not a superconductor, but on correlation $T_c^{Ta,RSh}(J_{RSh})$ of Eq. (5) it must be superconductor with $T_c$ = 7.4K. Here and further the number of compound in Table I is parenthetically shown. The calculation of $T_c$ for $TaB_2$ (N 29 [29, 44] and N30 [54]) with increasing parameter *c* points to the absence of superconductivity by both correlations.

Correlations $T_c^{Zr,RSh}(J_{RSh})$ and $T_c^{Zr,RP}(J_{RP})$ of Eqs. (7) and (8) built with of using of data on $ZrB_2$ [39], probably, have no physical sense, as fare as it is impossible to explain a sudden disappearance and appearance of superconductivity. Besides, $T_c$ calculated on the base of these correlations disagrees to experimental data this work [39], as far as indicates not on the absence but opposite on the presence of superconductivity with unrealistic high $T_c$ for diborides Nb, Ta and W, V, Hf, Pu, U also.

Thereby, a estimation has shown that the correlations of $T_c^{Ta,RSh}(J_{RSh})$ and $T_c^{Ta.RP}(J_{RP})$ are the most reliable. However, its can show only a trend $T_c$ change with changing of crystal chemical parameters of diborides, as for its building there was too little experimental data. The $T_c$ calculated on these correlations are referred to the diborides of stoichiometric composition, since there was no data for the account of non-stoichiometry of compounds by *J* calculation, while the last studies [16, 63, 64] point to Mg-deficiency and defects even in $MgB_2$. As a result, calculated $T_c^{Ta,RSh}$ of $MgB_2$ are little below than $T_c$ found experimentally ($T_c^{\exp}$). However, deficit of Mg (0.8, 0.9 and 1%) in samples, N54, 59 and 53 (Table), accordingly, raises $T_c^{Ta,RSh}$ (38.1 – 38.3 K) calculated on the Eq. (5) to the experimental values (38.8 – 39 K).

## 3. RESULTS AND DISCUSSION

Maximal $T_c$ values for diborides, calculated on correlation of $T_c^{Ta,RSh}(J_{RSh})$ and $T_c^{Ta.RP}(J_{RP})$ (Eqs. (5) and (6)) are 107 K at optimal value $J=J_0$ (by $J_0$ = -1.304) and 290 K (by $J_0$ = -3.386), accordingly. Increasing $T_c$ of diborides $AB_2$ by nearing $J$ to $J_0$ (Fig. 2 a) on the left occurs by shortening the distances $d$(B-B) in $B_2$ plane and "effective" interplanar distances $D_1$ and $D_2$. (As on the given interval $D_1$ and $D_2$ have negative values the modules $D_1$ and $D_2$ must increase). This is possible from shortening the lattice parameters $a$ and $c$, reducing a size of A atom and increasing its charge. Moreover, the most effect is reached when a charge of atom is increasing. So, WB2 (N5) has the most high $T_c$ (99.7 K and 290 K according to correlations $T_c^{Ta,RSh}(J_{RSh})$ and $T_c^{Ta.RP}(J_{RP})$) amongst considered diborides with $J<J_0$.

Besides, to rise $T_c$ of diborides with $J<J_0$ is possible by increasing $S$ coefficient to the account of introduction of vacancies in A layer or partial substitution A on the ions of smaller size or insignificantly differing from the size, but having smaller charge. For instance, $T_c$ of $WB_2$ and $MoB_2$ raise is possible by doping of W and Mo planes with the ions $Ru^{4+}$, $Os^{4+}$, $V^{5+}$, $Ti^{4+}$ (N18), $Nb^{5+}$ (N17) or $Al^{3+}$, but for $TaB_2$ by doping of Ta plane with $Al^{3+}$ (N34), $Ti^{4+}$ (N33) or $V^{5+}$( N31) ions. For arising a superconductivity in $NbB_2$ it is necessary to introduce Nb-vacancies or substitute part of $Nb^{5+}$ on $Al^{3+}$ (N41), $Ti^{4+}$ or $V^{5+}$ (N42). The $T_c$ of diborides with $J<J_0$ must increase under the action of pressure, unlike $MgB_2$, where $J>J_0$.

However, superconductors with $J<J_0$, having high $T_c$, are hitherto not discovered. As it is mentioned above, $TaB_2$ is a superconductor at $T_c$ = 9.5 K [35]. Cooper at al. [65] reported also, that in "boron-rich" $NbB_2$ compounds a superconductivity appears at $T_c$ = 3.87 K, and in $Zr_{0.13}Mo_{0.87}B_2$ at $T_c$ = 11 K. To our regret, the work [65] is inaccessible for us, but if expect that by substituting Mo on 13% Zr the lattice parameters $a$ and $c$ enlarged to 3.08 Å and 3.32 Å, accordingly, and there are no vacancies in $B_2$ plane, $T_c$ of $Zr_{0.13}Mo_{0.87}B_2$, calculated on Eq. (6) and Eq. (5) are 11 K and 86 K, accordingly.

Probably, a boron deficit in plane $B_2$ is a reason of absence of superconductivity in the diborides with $J<J_0$. It is possible a compression of $a$ parameter caused by reducing a size of «cation» A can rezult in arising the vacancies in plane $B_2$. Consequently, a decrease $T_c$ of compound appears up to full suppress its superconductivity. On the contrary, substitution in $MoB_2$ of part of $Mo^{6+}$ ions on large $Zr^{4+}$ ions with lower charge allows to obtain "boron-rich" diboride $Zr_{0.13}Mo_{0.87}B_2$ that is a superconducting properties. A problem appears to conservate a stoichiometry in $B_2$ plane in the diborides $AB_2$ with $J<J_0$ for the achievement of high $T_c$. May be, this can be reached only by partial substituting A on more large "cations" with the lower charge. Such substitution reduces $T_c$, that takes place in initial diborides but allows to obtain the superconductors with sufficiently high $T_c$. Dopants saving stoichiometry on boron for $AB_2$ (A = W, Mo, Ru or Os) can be the following cations: $Pb^{2+}$, $Mg^{2+}$, $Ag^{2+}$, $Sc^{3+}$, $Y^{3+}$, $Zr^{4+}$, $Sn^{4+}$, $Pb^{4+}$, $U^{4+}$ and $Th^{4+}$.

In diborides with $J>J_0$, unlike diborides with $J<J_0$, $T_c$ increase, when $J$ nears to $J_0$ on the right, occurs not with reducing but with increasing "effective" interplanar distances $D_1$ and $D_2$. By this as well as in the diborides with $J<J_0$ the $d$(B-B) distances must decrease or its increase is more slow than $D_1$ and $D_2$ raising. This can be reached by increasing $c$ lattice parameter with heightening the sizes of main or doping atoms of A plane and/or a $S$ coefficient to the account of introducing the vacancies in A plane or partial substitution A on the ions of greater size or ones insignificantly differing from the size but with another charge. Experimentally proved [7, 10-13] that substituting Al on Mg in $AlB_2$ results in arising superconductivity and increase $T_c$ to 39K in $Al_{1-x}Mg_xB_2$ with x growing. This substituting is accompanied with reducing $J$ by in overtaking growing of interplanar distances $D_1$and $D_2$ (N 80-88).

Usually inverse processes are considered, i.e. the suppress of superconductivity in $MgB_2$ by substituting Mg on Al or by pressing. Using the structural parameters from works [7, 17, 18, 40] we calculated the change of $T_c$ of $MgB_2$ from pressure (N70-79) and $Mg_{1-x}Al_xB_2$ from $x$ parameter (N 80-88). In both events $J$ increasing, accompanied by $T_c$ falling, occurs by greater reducing $c$ parameter in contrast to the parameter $a$. However, the linear reduction $T_c$ calculated on equations (5) and (6) with increase $P$ or x vastly below found in the experiment [5, 7, 10-23]. It is possible, that in this case as in the diborides with $J<J_0$, a reason of increasing a velocity of $T_c$ falling is raising a loss of boron atoms by shortening $a$ parameter with the growing of pressure or $x$ increase.

It is shown by experiment that isovalent substitution of $Mg^{2+}$ on $Zn^{2+}$ which size is only little more ($RSh$ =0 .88 Å, $RP$ = 0.74 Å) than Mg ($RSh$ = 0.86 Å, $RP$ = 0.65 Å) can very small to rise $T_c$ [9]. Our calculations (N 96-100) also confirm this conclusion. In work [7] it is theoretically predicted that partial substituting in $MgB_2$ of $Mg^{2+}$ on $Ca^{2+}$ or $Na^{1+}$ must result in the growing $T_c$ up to 52 K or 53 K, accordingly. Calculated by us structural parameters $T_c$ of these systems are close to the data of [7]: with the growing x from 0 to 0.2 $T_c^{Ta,RSh}$ ($T_c^{Ta.RP}$) of the systems $Mg_{(1-x)}Ca_xB_2$ (N 85, 89-92) and $Mg_{(1-x)}Na_xB_2$ (N 85, 93-95) goes up to 47 K (59 K) and 50 K (57 K), accordingly.

It follows from this that $T_c$ raising in $MgB_2$ can be reached by conservation of stoichiometry in $B_2$ plane by means of increasing the effective interplanar distances $D_1$ and $D_2$ by partial substituting $Mg^{2+}$ on the ions of greater size with charges 1+ or 2+, such as $Na^{1+}$, $K^{1+}$, $Cd^{2+}$ or $Ca^{2+}$.

It that way, the empirical dependence of $T_c(J)$ can be useful for the prognostication of the composition of new diborides with high $T_c$, and also for the estimation of $T_c$ and the correctness of determination of the structure $AB_2$ and the composition of A-cation planes. The $J$ value is not a simple ratio of geometric structural parameters of diborides. As in HTSC cuprates [27] a value of $J$ ratio depends on all these factors which influence on $T_c$ found experimentally, such as $d$(B-B) and $d$($B_2$-A), which have an original sense and give also an information about the hole concentration, and on the size and charge of A «cations" and doping atoms too.

## 4. CONCLUSIONS

In our work we examined a possibility of increasing $T_c$ in $AB_2$ diborides with the structure of $AlB_2$ on the ground of empirical correlation of $T_c$ with the crystal chemical parameters of anisotropic three-dimensional fragment – $A_2(B_2)$–sandwich, as there is in HTSC cuprates. By this parabolic dependence, $T_c$ is correlated with the ratio ($J$) of $d$(B-B) distances between B atoms situated at the ends of diagonal of hexagons in $B_2$ plane and sum of "effective" distances ($D_1+D_2$) from $B_2$ plane to two adjacent planes of A "cations", taking into account by calculation a charge size of these cations and doping atoms ($J = d(\text{B-B})/(D_1+D_2)$). We calculate the $T_c$ of diborides by this correlation (Table I). It follows:

- Among the diborides considered a superconductivity can to be only in diborides W, Mo, Ru, Os and Ta, where $J<J_0$, and Mg, Cu(II), Ag(I) and Au(I), where $J>J_0$. The result obtained there suggest that the empirical absence or low-temperature superconductivity established in transition metal diborides with $J<J_0$ might be explained of presence B vacancies in $B_2$ plane. In the absence of vacancies in $B_2$ plane $T_c$ of $AB_2$ diborides (A=W, Mo, Ru, Os) can be higher 77K, and in $TaB_2$ to reach 10K. For appearance of superconductivity and increasing $T_c$ in $NbB_2$ and $TaB_2$ to need introduction of Nb(Ta)-vacancies or partial substitution of $Nb^{5+}$ ($Ta^{5+}$) on $Al^{3+}$, $Ti^{4+}$ or $V^{5+}$.

- Partial substitution of W, Mo, Ru and Os on more large cations with lower charge decreases $T_c$ with respect to one in initial diborides, but allows to conservate a stoichiometry in $B_2$ plane and to make a superconductors with enough high $T_c$.
- Partial substitution in $MgB_2$ of Mg on larger but with lower charge "cations" must heighten $T_c$.
- By the pressure $T_c$ must increase in superconductors with $J<J_0$ and decrease in ones with $J>J_0$.
- Critical crystal chemical parameters controlling $T_c$, apart from the concentration of charge carriers in $B_2$ plane, are the distances between boron atoms in $B_2$ plane and the parameters characterised the space between of $B_2$ plane and A "cation" planes in sandwich $A_2(B_2)$, such as: an interval between the surface of the planes, the inhomogeneity surface of A-cation planes, and also the electric fields induced by the A "cations" and doping "cations" charges.

**ACKNOWLEDGMENTS**

This work was supported by the Russian Foundation for Basic research under grant 00-03-32486.

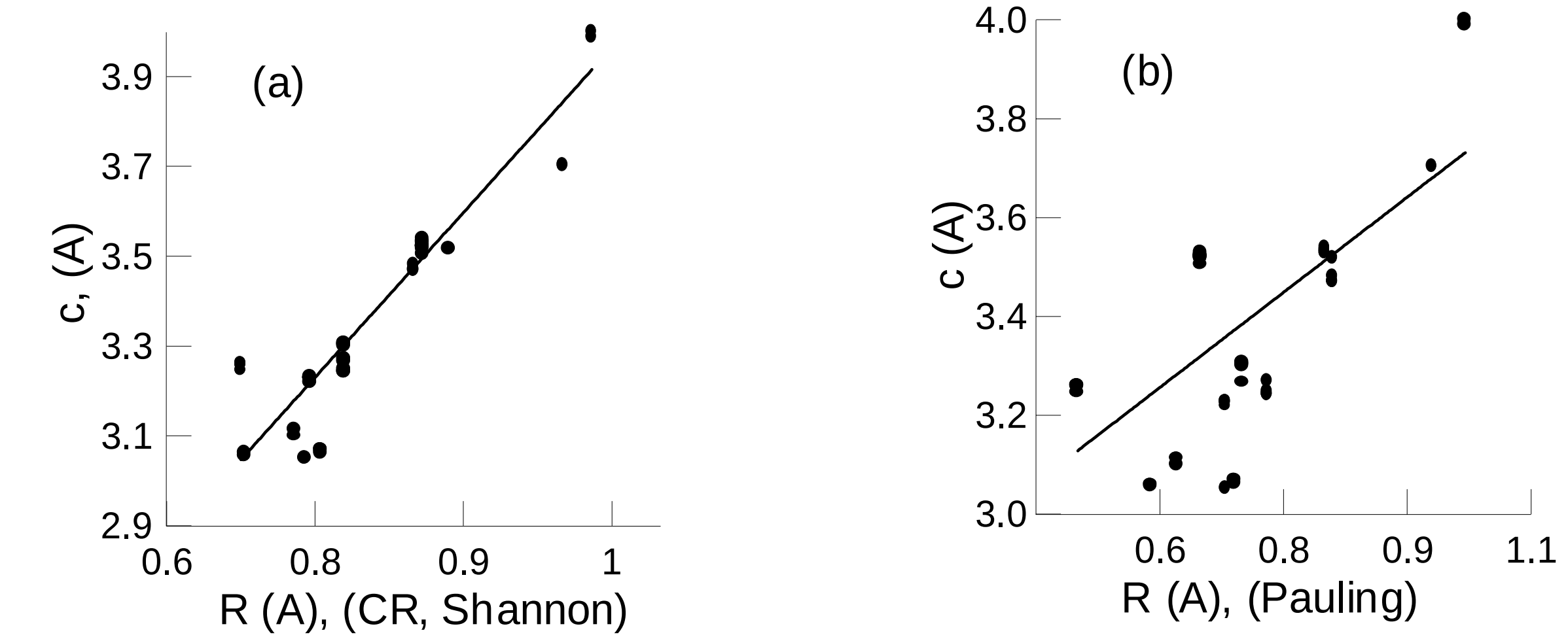

**Fig. 1.** The variation of the lattice parameters $c$ in $AB_2$ diborides as a function of Shannon crystal radii (a) and Pauling crystal ion radii (b) of A “cations”.

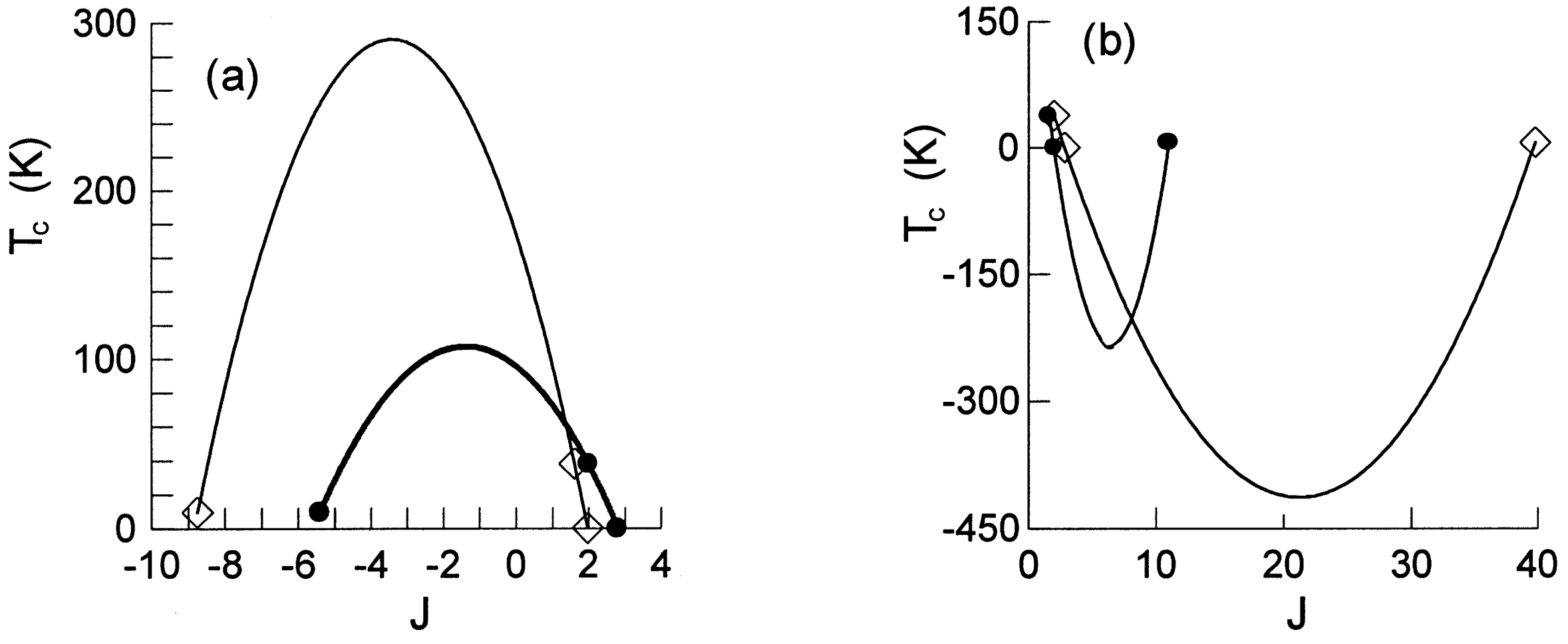

**Fig. 2.** $T_c$ as a function of J in the diborides $AB_2$. $J$ was calculated by using Shannon (solid symbols -•) and Pauling (open symbols - ) radii systems: (a) $T_c^{Ta,RSh}(J_{RSh})$ and $T_c^{Ta.RP}(J_{RP})$ of Eq. 5 and Eq.6 obtained on the data for $MgB_2$, $AlB_2$ and $TaB_2$; (b) $T_c^{Zr,RSh}(J_{RSh})$ and $T_c^{Zr,RP}(J_{RP})$ of Eq. 7 and Eq.8 obtained on the data for $MgB_2$, $AlB_2$ and $ZrB_2$.

TABLE I. The estimated $T_c$ ($T_c^{Ta,RSh}$ and $T_c^{Ta,RP}$) and experimental lattice parameters of diborides $AB_2$ used for calculation.

| N | Compound | $T_c^{exp}$ (K) | $a$ (Å) | $c$ (Å) | $D_1^{RSh} + D_2^{RSh}$ | $J_{RSh}$ | $T_c^{Ta,RSh}$ (K) | $D_1^{RP} + D_2^{RP}$ | $J_{RP}$ | $T_c^{Ta,RP}$ (K) | Refer. |
|---|---|---|---|---|---|---|---|---|---|---|---|
| 1. | $RuB_2$ (6+) | - | 2.852 | 2.855 | -1.255 | -2.624 | 98.1 | - | - | - | [29] |
| 2. | $OsB_2$ (6+) | - | 2.876 | 2.871 | -1.239 | -2.680 | 97.2 | - | - | - | [29] |
| 3. | $MnB_2$ (4+) | - | 3.007 | 3.037 | 0.357 | 9.726 | Non-SC | - | - | | [29], [62] |
| 4. | $Mn_{0.64}Mo_{0.36}B_2$ | - | 3.036 | 3.098 | 0.513 | 6.828 | Non-SC | - | - | | [41] |
| 5. | $WB_2$ | - | 3.020 | 3.050 | -1.390 | -2.509 | 99.7 | -1.030 | -3.386 | 290.0 | [42] |
| 6. | $VB_2$ | - | 2.998 | 3.057 | -0.343 | 10.093 | Non-SC | 0.107 | 32.353 | Non-SC | [42], [61] |
| 7. | $VB_2$ | - | 3.000 | 3.060 | -0.340 | 10.188 | Non-SC | 0.110 | 31.492 | Non-SC | [43] |
| 8. | $V_{0.50}Cr_{0.50}B_2$ | - | 2.990 | 3.045 | -0.426 | -8.102 | Non-SC | 0.112 | 30.917 | Non-SC | [44] |
| 9. | $CrB_2$ (3+) | - | 2.969 | 3.0668 | 0.801 | 4.280 | Non-SC | 0.996 | 3.442 | Non-SC | [45] |
| 10. | $CrB_2$ (3+) | - | 2.970 | 3.070 | 0.805 | 4.260 | Non-SC | 1.000 | 3.429 | Non-SC | [29] |
| 11. | $Cr_{0.85}Al_{0.15}B_2$ | - | 2.992 | 3.106 | 0.851 | 4.057 | Non-SC | 1.081 | 3.197 | Non-SC | [46] |
| 12. | $Cr_{0.60}Mo_{0.40}B_2$ | - | 3.069 | 3.112 | 1.163 | 3.046 | Non-SC | 0.710 | 4.992 | Non-SC | [44] |
| 13. | $MoB_2$ | - | 3.039 | 3.055 | -1.325 | -2.648 | 97.7 | -0.665 | -5.277 | 256.2 | [47] |
| 14. | $MoB_2$ | - | 3.050 | 3.080 | -1.300 | -2.709 | 96.8 | -0.640 | -5.503 | 247.4 | [44] |
| 15. | $MoB_2$ | - | 3.050 | 3.113 | -1.267 | -2.780 | 95.6 | -0.607 | -5.802 | 234.2 | [29], [48] |
| 16. | $MoB_2$ | - | 3.005 | 3.173 | -1.207 | -2.875 | 94.0 | -0.547 | -6.344 | 205.7 | [49] |
| 17. | $Mo_{0.72}Nb_{0.28}B_2$ | - | 3.068 | 3.143 | -1.276 | -2.776 | 95.7 | -0.587 | -6.039 | 222.5 | [50] |
| 18. | $Mo_{0.50}Ti_{0.50}B_2$ | - | 3.044 | 3.207 | -1.396 | -2.518 | 99.6 | -0.593 | -5.931 | 227.9 | [44] |
| 19. | $TiB_2$ | - | 3.030 | 3.227 | 0.247 | 14.165 | Non-SC | 0.507 | 6.901 | Non-SC | [45] |
| 20. | $TiB_2$ | - | 3.030 | 3.230 | 0.250 | 13.995 | Non-SC | 0.510 | 6.860 | Non-SC | [29] |
| 21. | $TiB_2$ | - | 3.031 | 3.238 | 0.258 | 13.560 | Non-SC | 0.518 | 6.757 | Non-SC | [51] |
| 22. | $TiB_2$ | - | 3.038 | 3.239 | 0.259 | 13.540 | Non-SC | 0.519 | 6.759 | Non-SC | [52] |
| 23. | $Ti_{0.50}Cr_{0.50}B_2$ | - | 2.990 | 3.140 | 0.182 | 19.000 | Non-SC | 0.477 | 7.239 | Non-SC | [44] |
| 24. | $CuB_2$ (2+) | 65* | 2.960* | 3.250* | 1.510 | 2.263 | 26.0 | 1.870 | 1.828 | 15.3 | [53] |
| 25. | $AgB_2$ (3+) | - | 3.000 | 3.240 | 0.570 | 6.077 | Non-SC | - | - | Non-SC | [29] |

TABLE I. (cont.)

| N | Compound | $T_c^{\exp}$ (K) | $a$ (Å) | $c$ (Å) | $D_1^{RSh}+D_2^{RSh}$ | $J_{RSh}$ | $T_c^{Ta,RSh}$ (K) | $D_1^{RP}+D_2^{RP}$ | $J_{RP}$ | $T_c^{Ta,RP}$ (K) | Refer. |
|---|---|---|---|---|---|---|---|---|---|---|---|
| 26. | $AgB_2$ (1+) | 59* | 2.980* | 3.920* | 2.630 | 1.308 | 62.8 | 2.660 | 1.294 | 68.3 | [53] |
| 27. | $TaB_2$ | 9.5 | 3.082 | 3.243 | -0.657 | -5.417 | 9.5 | -0.407 | -8.744 | 9.5 | [37] |
| 28. | $TaB_2$ | Non-SC | 3.087 | 3.247 | -0.6538 | -5.459 | 7.4 | -0.403 | -8.845 | Non-SC | [39] |
| 29. | $TaB_2$ | - | 3.080 | 3.270 | -0.630 | -5.645 | Non-SC | -0.380 | -9.359 | Non-SC | [29], [44] |
| 30. | $TaB_2$ | - | 3.065 | 3.283 | -0.617 | -5.736 | Non-SC | -0.367 | -9.643 | Non-SC | [54] |
| 31. | $Ta_{0.50}V_{0.50}B_2$ | - | 3.040 | 3.160 | -0.791 | -4.450 | 50.8 | -0.542 | -6.476 | 197.9 | [55] |
| 32. | $Ta_{0.50}Cr_{0.50}B_2$ | - | 3.025 | 3.210 | -0.873 | -4.000 | 66.2 | -0.561 | -6.220 | 212.7 | [44] |
| 33. | $Ta_{0.50}Ti_{0.50}B_2$ | - | 3.050 | 3.246 | -0.741 | -4.750 | 39.2 | -0.463 | -7.607 | 116.7 | [44] |
| 34. | $Ta_{0.77}Al_{0.23}B_2$ | - | 3.060 | 3.294 | -0.681 | -5.185 | 20.4 | -0.412 | -8.580 | 26.6 | [56] |
| 35. | $Ta_{0.50}Hf_{0.50}B_2$ | - | 3.110 | 3.370 | -0.566 | -6.341 | Non-SC | -0.297 | -12.105 | Non-SC | [55] |
| 36. | $Ta_{0.50}Zr_{0.50}B_2$ | - | 3.120 | 3.400 | -0.531 | -6.780 | Non-SC | -0.266 | -12.762 | Non-SC | [44] |
| 37. | $AlB_2$ | - | 3.005 | 3.257 | 1.232 | 2.816 | Non-SC | 1.757 | 1.975 | Non-SC | [40] |
| 38. | $AlB_2$ | - | 3.009 | 3.262 | 1.237 | 2.809 | 0 | 1.762 | 1.972 | 0 | [29], [57] |
| 39. | $NbB_2$ | Non-SC | 3.110 | 3.267 | -0.633 | -5.673 | Non-SC | -0.233 | -15.412 | Non-SC | [39] |
| 40. | $NbB_2$ | - | 3.090 | 3.300 | -0.600 | -5.947 | Non-SC | -0.200 | -17.840 | Non-SC | [29] |
| 41. | $Nb_{0.67}Al_{0.33}B_2$ | - | 3.068 | 3.334 | -0.673 | -5.267 | 16.7 | -0.205 | -17.319 | Non-SC | [56] |
| 42. | $Nb_{0.50}V_{0.50}B_2$ | - | 3,030 | 3.200 | -0.748 | -4.678 | 42.1 | -0.326 | -10.747 | Non-SC | [55] |
| 43. | $Nb_{0.50}Zr_{0.50}B_2$ | - | 3.128 | 3.420 | 0.510 | -7.081 | Non-SC | -0.084 | -43.220 | Non-SC | [44] |
| 44. | $HfB_2$ | - | 3.140 | 3.470 | 0.070 | 51.797 | Non-SC | 0.230 | 15.764 | Non-SC | [29] |
| 45. | $HfB_2$ | - | 3.141 | 3.470 | 0.070 | 51.813 | Non-SC | 0.230 | 15.769 | Non-SC | [42], [44] |
| 46. | $HfB_2$ | - | 3.139 | 3.473 | 0.073 | 49.650 | Non-SC | 0.233 | 15.558 | Non-SC | [58] |
| 47. | $Hf_{0.50}Ti_{0.50}B_2$ | - | 3.085 | 3.368 | -0.034 | -104.4 | Non-SC | 0.139 | 25.595 | Non-SC | [44] |
| 48. | $AuB_2$ (3+) | - | 3.140 | 3.510 | 0.540 | 6.714 | Non-SC | - | - | - | [29] |
| 49. | $AuB_2$ (1+) | 72* | 2.980* | 4.050* | 2.540 | 1.355 | 61.3 | 2.680 | 1.284 | 69.2 | [53] |
| 50. | $ScB_2$ | - | 3.146 | 3.517 | 0.863 | 4.209 | Non-SC | 1.087 | 3.342 | Non-SC | [29] |
| 51. | $MgB_2$ | 49.0 | 3.068 | 3.505 | 1.785 | 1.985 | 37.8 | 2.205 | 1.607 | 37.9 | [31] |

TABLE I. (cont.)

| N | Compound | $T_c^{exp}$ (K) | $a$ (Å) | $c$ (Å) | $D_1^{RSh}+D_2^{RSh}$ | $J_{RSh}$ | $T_c^{Ta,RSh}$ (K) | $D_1^{RP}+D_2^{RP}$ | $J_{RP}$ | $T_c^{Ta,RP}$ (K) | Refer. |
|---|---|---|---|---|---|---|---|---|---|---|---|
| 52. | $MgB_2$ | 36.6 | 3.075 | 3.519 | 1.799 | 1.974 | 38.3 | 2.219 | 1.600 | 38.6 | [32] |
| 53. | $MgB_2$ | 39.0 | 3.0856 | 3.5199 | 1.800 | 1.979 | 38.1 | 2.220 | 1.605 | 38.1 | [33] |
| 54. | $MgB_2$ | 38.8 | 3.083 | 3.520 | 1.800 | 1.978 | 38.1 | 2.220 | 1.604 | 38.2 | [34] |
| 55. | $MgB_2$ | - | 3.085 | 3.520 | 1.800 | 1.979 | 38.1 | 2.220 | 1.605 | 38.1 | [7] |
| 56. | $MgB_2$ | 38.1 | 3.0851 | 3.5201 | 1.800 | 1.979 | 38.1 | 2.220 | 1.605 | 38.1 | [36] |
| 57. | $MgB_2$ | 39.0 | 3.0849 | 3.5211 | 1.801 | 1.978 | 38.1 | 2.221 | 1.604 | 38.2 | [35] |
| 58. | $MgB_2$ | 38.9 | 3.0846 | 3.5230 | 1.803 | 1.975 | 38.3 | 2.223 | 1.602 | 38.4 | [5] |
| 59. | $MgB_2$ | 39.0 | 3.086 | 3.524 | 1.804 | 1.975 | 38.3 | 2.224 | 1.602 | 38.3 | [1] |
| 60. | $ZrB_2$ | - | 3.169 | 3.523 | 0.083 | 44.084 | Non-SC | 0.323 | 11.328 | Non-SC | [59] |
| 61. | ZrB2 | - | 3.150 | 3.530 | 0.090 | 40.415 | Non-SC | 0.330 | 11.022 | Non-SC | [29] |
| 62. | ZrB2 | 5.5 | 3.170 | 3.532 | 0.092 | 39.787 | Non-SC | 0.332 | 11.025 | Non-SC | [39] |
| 63. | ZrB2 | - | 3.166 | 3.535 | 0.095 | 38.482 | Non-SC | 0.335 | 10.913 | Non-SC | [45] |
| 64. | $Zr_{0.50}Ti_{0.50}B_2$ | - | 3.098 | 3.390 | -0.054 | -66.76 | Non-SC | 0.205 | 17.414 | Non-SC | [44] |
| 65. | $LuB_2$ | - | 3.246 | 3.704 | 0.702 | 5.339 | Non-SC | 0.914 | 4.101 | Non-SC | [42] |
| 66. | $YB_2$ | - | 3.290 | 3.835 | 0.715 | 5.313 | Non-SC | 1.045 | 3.635 | Non-SC | [60] |
| 67. | $PuB^2$ (4+) | - | 3.180 | 3.900 | -0.100 | -36.72 | Non-SC | - | - | - | [29] |
| 68. | $UB_2$ (4+) | - | 3.136 | 3.988 | -0.132 | 27.433 | Non-SC | 0.108 | 33.529 | Non-SC | [42] |
| 69. | $UB_2$ (4+) | - | 3.140 | 4.000 | -0.120 | 30.215 | Non-SC | 0.120 | 30.215 | Non-SC | [29] |
| **70.** | **$MgB_2$** | **38.2** | **3.0859** | **3.5212** | **1.801** | **1.978** | **38.1** | **2.221** | **1.604** | **38.2** | [17] |
| 71. | $MgB_2$, 1.17 GPa | - | 3.0802 | 3.5112 | 1.791 | 1.986 | 37.8 | 2.211 | 1.608 | 37.8 | [17] |
| 72. | $MgB_2$, 2.14 GPa | - | 3.0715 | 3.4985 | 1.778 | 1.994 | 37.5 | 2.198 | 1.613 | 37.3 | [17] |
| 73. | $MgB_2$, 3.05 GPa | - | 3.0671 | 3.4885 | 1.768 | 2.002 | 37.2 | 2.188 | 1.618 | 36.7 | [17] |
| 74. | $MgB_2$, 4.07 GPa | - | 3.0635 | 3.4819 | 1.762 | 2.008 | 36.9 | 2.182 | 1.621 | 36.4 | [17] |
| 75. | $MgB_2$, 5.09 GPa | - | 3.0545 | 3.4718 | 1.752 | 2.013 | 36.7 | 2.172 | 1.624 | 36.1 | [17] |
| 76. | $MgB_2$, 6.53 GPa | - | 3.0497 | 3.4586 | 1.739 | 2.025 | 36.2 | 2.159 | 1.631 | 35.4 | [17] |
| 77. | $MgB_2$, 8.02 GPa | - | 3.0484 | 3.4572 | 1.737 | 2.026 | 36.2 | 2.157 | 1.632 | 35.3 | [17] |

TABLE I. (cont.)

| N | Compound | $T_c^{exp}$ (K) | $a$ (Å) | $c$ (Å) | $D_1^{RSh}+D_2^{RSh}$ | $J_{RSh}$ | $T_c^{Ta,RSh}$ (K) | $D_1^{RP}+D_2^{RP}$ | $J_{RP}$ | $T_c^{Ta,RP}$ (K) | Refer. |
|---|---|---|---|---|---|---|---|---|---|---|---|
| **78.** | **$MgB_2$** | **-** | **3.0906** | **3.5287** | **1.809** | **1.973** | **38.3** | **2.229** | **1.601** | **38.5** | [18] |
| 79. | $MgB_2$, 6.15 GPa | - | 3.0646 | 3.4860 | 1.766 | 2.004 | 37.1 | 2.186 | 1.619 | 36.6 | [18] |
| **80.** | **$MgB_2$** | **-** | **3.085** | **3.523** | **1.803** | **1.976** | **38.2** | **2.223** | **1.602** | **38.4** | [40] |
| 81. | $Al_{0.39}Mg_{0.61}B_2$ | - | 3.047 | 3.369 | 1.763 | 1.996 | 37.4 | 2.193 | 1.604 | 38.1 | [40] |
| 82. | $Al_{0.50}Mg_{0.50}B_2$ | - | 3.047 | 3.366 | 1.450 | 2.426 | 18.6 | 2.000 | 1.759 | 22.4 | [40] |
| 83. | $Al_{0.67}Mg_{0.33}B_2$ | - | 3.037 | 3.331 | 1.374 | 2.552 | 12.8 | 1.915 | 1.831 | 14.9 | [40] |
| 84. | $Al_{0.75}Mg_{0.25}B_2$ | - | 3.030 | 3.302 | 1.327 | 2.637 | 8.6 | 1.864 | 1.877 | 10.1 | [40] |
| **85.** | **$MgB_2$** | **-** | **3.065** | **3.5186** | **1.799** | **1.968** | **38.5** | **2.219** | **1.595** | **39.1** | [7] |
| 86. | $Mg_{0.98}Al_{0.02}B_2$ | - | 3.084 | 3.5158 | 1.802 | 1.976 | 38.2 | 2.223 | 1.602 | 38.4 | [7] |
| 87. | $Mg_{0.96}Al_{0.04}B_2$ | - | 3.083 | 3.5115 | 1.804 | 1.974 | 38.3 | 2.225 | 1.600 | 38.6 | [7] |
| 88. | $Mg_{0.92}Al_{0.08}B_2$ | - | 3.081 | 3.4969 | 1.802 | 1.974 | 38.3 | 2.224 | 1.600 | 38.6 | [7] |
| 89. | $Mg_{0.95}Ca_{0.05}B_2$ | - | 3.072* | 3.5451* | 1.855 | 1.912 | 40.8 | 2.304 | 1.541 | 44.4 | [7] |
| 90. | $Mg_{0.90}Ca_{0.10}B_2$ | - | 3.080* | 3.5728* | 1.913 | 1.859 | 42.9 | 2.392 | 1.487 | 49.7 | [7] |
| 91. | $Mg_{0.85}Ca_{0.15}B_2$ | - | 3.087* | 3.6025* | 1.974 | 1.805 | 45.0 | 2.483 | 1.435 | 54.8 | [7] |
| 92. | $Mg_{0.80}Ca_{0.20}B_2$ | 52* | 3.095* | 3.6304* | 2.035 | 1.756 | 46.9 | 2.574 | 1.388 | 59.3 | [7] |
| 93. | $Mg_{0.95}Na_{0.05}B_2$ | - | 3.063* | 3.5592* | 1.870 | 1.892 | 41.6 | 2.288 | 1.546 | 43.9 | [7] |
| 94. | $Mg_{0.90}Na_{0.10}B_2$ | - | 3.061* | 3.5967* | 1.940 | 1.822 | 44.4 | 2.356 | 1.500 | 48.4 | [7] |
| 95. | $Mg_{0.80}Na_{0.20}B_2$ | 53* | 3.057* | 3.6776* | 2.094 | 1.686 | 49.6 | 2.505 | 1.409 | 57.3 | [7] |
| **96.** | **MgB2** | **38.5** | **3.0787** | **3.5178** | **1.798** | **1.977** | **38.2** | **2.218** | **1.603** | **38.3** | [8] |
| 97. | $Mg_{0.97}Zn_{0.03}B_2$ | 38.4 | 3.0870 | 3.5241 | 1.805 | 1.974 | 38.3 | 2.233 | 1.5961 | 38.9 | [9] |
| 98. | $Mg_{0.95}Zn_{0.05}B_2$ | 38 | 3.0803 | 3.5226 | 1.805 | 1.972 | 38.4 | 2.238 | 1.589 | 39.7 | [8] |
| 99. | $Mg_{0.90}Zn_{0.10}B_2$ | 38.3 | 3.0841 | 3.5250 | 1.809 | 1.968 | 38.5 | 2.256 | 1.579 | 40.6 | [8] |
| 100. | $Mg_{0.80}Zn_{0.20}B_2$ | 38.3 | 3.0841 | 3.5239 | 1.812 | 1.965 | 38.7 | 2.285 | 1.558 | 42.7 | [8] |

* - $T_c$ and lattice constant are calculate.